# Electromagnetic Waves Reflectance of Graphene – Magnetic Semiconductor Superlattice in Magnetic Field

Dmitry A. Kuzmin, *Member, IEEE*, Igor V. Bychkov, *Member, IEEE,* and Vladimir G. Shavrov

*Abstract*— Electrodynamic properties of the graphene – magnetic semiconductor – graphene superlattice placed in magnetic field have been investigated theoretically in Faraday geometry with taking into account dissipation processes. Frequency and field dependences of the reflectance, transmittance and absorbtance of electromagnetic waves by such superlattice have been calculated for different numbers of periods of the structure and different sizes of the periods with using a transfer matrix method. The possibility of efficient control of electrodynamic properties of graphene – magnetic semiconductor – graphene superlattice has been shown.

## I. Introduction

Nowadays graphene, two-dimensional honeycomb-like lattice of carbon atoms, attracts researchers' attention with their special properties, including electronic and electrodynamic ones [1 - 4]: linear dispersion of carriers; room-temperature quantum Hall effect;. graphene layer can support highly localized surface electromagnetic waves— surface plasmon polaritons [5-8]; the waveguide modes with the negative group speed can exist in the structure of two graphene layers with a layer of dielectric [9]; the hyperbolic metamaterial based on graphene – dielectric multilayer structure may be created [10]; etc. Despite the large number of studies, the authors are usually limited themselves by investigation of a non-magnetic dielectric medium, where graphene is placed. Recently, the dispersion properties of an anisotropic metamaterial composed of periodic stacking of graphene-liquid crystal layers have been investigated [11]. The switching between the elliptic and hyperbolic dispersion phases via control of the temperature, voltage and external electric field has been studied. It has been shown that this switching can be used to control of the transmission and reflection at the interface of the metamaterial and air. So, it is very interesting to study the dynamic characteristics of graphene structures with more complex materials. A magnetic semiconductor could be an example of such material. Frequency dispersion of the permittivity is one of the significant differences of semiconductor from dielectric; the plasma waves can excite in the semiconductor structures. When the semiconductor is placed in an external magnetic field, the helicons can to propagate in the material. Their properties depend on the magnetic field value. In its turn, the magnetic semiconductors, for example, may have a large magnetoresistance, magnetooptical properties, etc. Thus, the electrodynamic properties of graphene-magnetic semiconductor-based structures can be quite interesting. This paper is devoted to investigation of the electrodynamical properties of graphene - magnetic semiconductor superlattice placed in an external magnetic field.

## II. Geometry of The Problem

Geometry of the problem is represented in Fig. 1. The structure is a periodic stacking of graphene monolayers and magnetic semiconductor layers. Suppose that linearly polarized plane electromagnetic wave is normally incident in the surface of the structure and that an external magnetic field **H₀** is directed perpendicular to the structure surface (i.e. Faraday geometry). Due to the axial symmetry of the problem it is sufficient to consider an electromagnetic wave polarized along only one axis. The coordinate axes are chosen so that the $z$ axis coincides with the direction of the external magnetic field. The thickness of the magnetic semiconductor is denoted

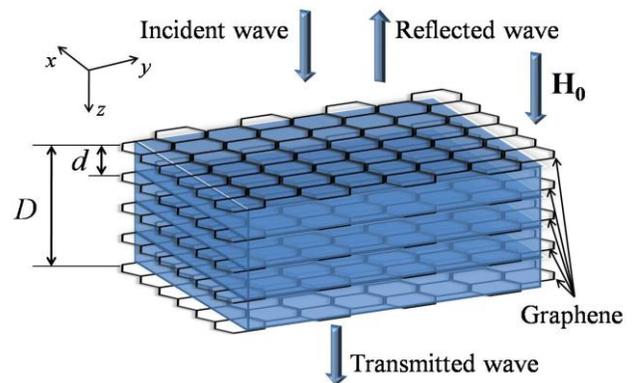

Fig. 1. (Color online) Geometry of the structure with the thickness *D* composed of a periodic structure of graphene layers separated by magnetic semiconductor slabs with the thickness *d* placed in external magnetic field **H₀**.

7 March 2014.
This work was supported in part by the RFBR under Grant # 13-07-00462.
D. A. Kuzmin is with the Chelyabinsk State University, Chelyabinsk 454001 Russia (e-mail: kuzminda89@gmail.com).
I. V. Bychkov is with the Chelyabinsk State University, Chelyabinsk 454001 Russia (e-mail: bychkov@csu.ru).
V. G. Shavrov is with the Kotelnikov Institute of Radioengineering and Electronics of RAS, Moscow 125009 Russia (e-mail: shavrov@cplire.ru).

$d$; $D$ is the thickness of all the structure.

For solving this problem, one has to know the characteristics of each component of the structure. For the magnetic semiconductor such characteristics are the tensors of the permeability $\hat{\mu}$ and the permittivity $\hat{\varepsilon}$. The permeability tensor of the magnetic semiconductor placed in an external magnetic field can be described as following:

$$\hat{\mu} = \begin{pmatrix} \mu_\perp & i\mu_a & 0 \\ -i\mu_a & \mu_\perp & 0 \\ 0 & 0 & \mu_\| \end{pmatrix};$$

$$\mu_\perp = 1 + \frac{\omega_M(\omega_H - i\alpha\omega)}{\omega_H^2 - (1+\alpha^2)\omega^2 - 2i\alpha\omega\omega_H}; \quad (1)$$

$$\mu_a = \frac{-\omega_M\omega}{\omega_H^2 - (1+\alpha^2)\omega^2 - 2i\alpha\omega\omega_H}; \quad \mu_\| = 1 - \frac{i\alpha\omega_M}{\omega + i\alpha\omega_H}.$$

In (1) we used the following notation: $\omega_H = gH_0$, $\omega_M = 4\pi gM_0$, $g$ is the gyromagnetic ratio, $M_0$ is the saturation magnetization, $\alpha$ is the damping parameter.

The permittivity tensor has the usual form:

$$\hat{\varepsilon} = \begin{pmatrix} \varepsilon_\perp & i\varepsilon_a & 0 \\ -i\varepsilon_a & \varepsilon_\perp & 0 \\ 0 & 0 & \varepsilon_\| \end{pmatrix}; \varepsilon_\perp = \varepsilon_0\left(1 - \frac{\omega_p^2(\omega+i\nu)}{\omega\left[(\omega+i\nu)^2 - \omega_c^2\right]}\right);$$

$$\varepsilon_a = \varepsilon_0 \frac{\omega_p^2\omega_c}{\omega\left[(\omega+i\nu)^2 - \omega_c^2\right]}; \varepsilon_\| = \varepsilon_0\left(1 - \frac{\omega_p^2}{\omega(\omega+i\nu)}\right). \quad (2)$$

Here, $\varepsilon_0$ is the lattice caused part of the permittivity, $\omega_p = \sqrt{4\pi n_s e^2/m^*}$ and $\omega_c = eH_0/m^*c$ are the plasma and the cyclotron frequencies, consequently, $e$ and $m^*$ are the charge and the effective mass of carriers, $n_s$ is the carriers density, $\nu$ is the effective collision rate.

Note, that for representation of magnetic semiconductor like a material with the tensors of the permeability (1) in the presented geometry, magnetization of semiconductor should be directed along an external magnetic field (perpendicular to the structure). Such scenario will take a place when an external magnetic field value is more than demagnetizing field, $H_0 > 4\pi M_0$; in formulas (1) and (2) in place of $H_0$ should be written an effective magnetic field value $H_0^{eff} = H_0 - 4\pi M_0$. We will take into account this fact, but will not to write superscript "eff" in further.

Graphene can be represented as a conductive surface [5] with the frequency dependent tensor of conductivity $\hat{\sigma}$, which components have been obtained in [12, 13]

$$\hat{\sigma} = \begin{pmatrix} \sigma_0 & \sigma_H \\ -\sigma_H & \sigma_0 \end{pmatrix};$$

$$\sigma_0 = \frac{e^2 v_F^2 |eH_0|(\hbar\omega + 2i\Gamma)}{i\pi c} \times$$

$$\times \sum_n \left\{ \begin{array}{l} \frac{[n_F(M_n) - n_F(M_{n+1})] - [n_F(-M_n) - n_F(-M_{n+1})]}{(M_{n+1} - M_n)^3 - (\hbar\omega + 2i\Gamma)^2(M_{n+1} - M_n)} + \\ + \frac{[n_F(-M_n) - n_F(M_{n+1})] - [n_F(M_n) - n_F(-M_{n+1})]}{(M_{n+1} + M_n)^3 - (\hbar\omega + 2i\Gamma)^2(M_{n+1} + M_n)} \end{array} \right\};$$

$$\sigma_H = -\frac{e^2 v_F^2 eH_0}{\pi c} \times$$

$$\times \sum_n \left\{ \begin{array}{l} \{[n_F(M_n) - n_F(M_{n+1})] + [n_F(-M_n) - n_F(-M_{n+1})]\} \times \\ \times \frac{2\left[M_n^2 + M_{n+1}^2 - (\hbar\omega + 2i\Gamma)^2\right]}{\left[M_n^2 + M_{n+1}^2 - (\hbar\omega + 2i\Gamma)^2\right]^2 - 4M_n^2 M_{n+1}^2} \end{array} \right\}$$

(3)

Here, $M_n = v_F(2neH_0/c)^{1/2}$ is energy of the corresponding Landau level, $n_F$ is the function of Fermi-Dirac distribution, $v_F$ is the velocity of electrons on the Fermi surface, $\Gamma/\hbar$ is the scattering rate.

For solving the problem one has to use the system of Maxwell's equations

$$rot\mathbf{E} = -c^{-1}\partial\mathbf{B}/\partial t; \quad rot\mathbf{H} = c^{-1}\partial\mathbf{D}/\partial t \quad (4)$$

with the material equations

$$\mathbf{D} = \hat{\varepsilon}\mathbf{E}; \quad \mathbf{B} = \hat{\mu}\mathbf{H}; \quad \mathbf{j} = \hat{\sigma}\mathbf{E} \quad (5)$$

and the boundary conditions

$$(\mathbf{E}_2 - \mathbf{E}_1) \times \mathbf{n}_{12} = 0; \quad (\mathbf{H}_2 - \mathbf{H}_1) \times \mathbf{n}_{12} = 4\pi\mathbf{j}/c \quad (6)$$

where, indexes 1 and 2 means the fields in the first and the second medium, $\mathbf{n}_{12}$ is the normal vector to the partition surface directed from the first medium to the second one, $\mathbf{j}$ is the density of the surface current in graphene layer.

Dispersion equation of the magnetic semiconductor has a usual for bigyrotropic medium form [14]:

$$k_\pm = k_0\sqrt{(\varepsilon_\perp \pm \varepsilon_a)(\mu_\perp \pm \mu_a)}, \quad (7)$$

where $k_0 = \omega/c$ is the wave number of electromagnetic wave in vacuum, indexes "+" and "-" correspond to the right- and left- polarized waves.

Solving the system of equations (4)-(6) with the dispersion equation (7) for each layer of magnetic semiconductor we obtain the amplitudes of reflected and transmitted waves. Then reflectance $R$ and transmittance $T$ can be found:

$$R = \frac{|E_{xR}|^2 + |E_{yR}|^2}{|E_{x0}|^2 + |E_{y0}|^2}; \quad T = \frac{|E_{xT}|^2 + |E_{yT}|^2}{|E_{x0}|^2 + |E_{y0}|^2}, \quad (8)$$

where indexes "$R$" and "$T$" denote amplitudes of reflected and transmitted waves, consequently, index "0" denotes amplitude of incident wave. Obtaining their values we define the absorptance $A = 1 - R - T$.

Such a method of solving the problem is not very convenient due to sharp increase in the number of equations with increasing in the periods of the structure. Using a transfer matrix method is more convenient for solving our problem. When the electromagnetic wave in the medium can be classified into TE- and TM- polarization, transfer matrixes 2x2 are usually used. In our case due to gyrotropy of the medium electromagnetic wave cannot to be classified into TE- and TM- polarization and one have to use a transfer matrixes 4x4. Transfer matrix $\hat{M}$ connects the amplitudes of the tangential components of the electric and magnetic fields before and after layer of the material:

$$\Xi_{before} = \hat{M}\Xi_{after}; \quad \Xi = \left(E_x, E_y, H_x, H_y\right)^T. \quad (9)$$

Will denote the transfer matrix of graphene layer $\hat{M}_g$ and the transfer matrix of magnetic semiconductor layer $\hat{M}_{ms}$. So, the transfer matrix of one period of the structure (graphene – magnetic semiconductor) is $\hat{M}_1 = \hat{M}_g \hat{M}_{ms}$ and the transfer matrix of all the structure is $\hat{M} = \left(\hat{M}_1\right)^N \hat{M}_g$, where $N = D/d$ is number of the periods. When the matrix $\hat{M}$ is known we have the following equations for obtaining the amplitudes of reflected and transmitted waves: $\Xi_0 + \Xi_R = \hat{M}\Xi_T$.

For the numerical simulation we will use the characteristic parameters of magnetic semiconductor $CdCr_2Se_4$ with Curie temperature $T_C = 130$ K [15, 16]:

$$M_0 = 350 \text{ G}, \alpha = 0.1, g = 1.75 \cdot 10^7 \text{ Oe}^{-1}\text{s}^{-1},$$
$$\varepsilon_0 = 20, m^* = 0.15 m_e, n_s = 10^{18} \text{cm}^{-3}, \nu = 10^{15} \text{s}^{-1}. \quad (10)$$

For simulation of the graphene properties will use parameters $v_F = 10^8$ cm/s, $\Gamma = 2 \cdot 10^{-15}$ erg, value of the chemical potential $\mu_{chem}$ from the Fermi-Dirac distribution function depend on temperature $T$ and carrier density in graphene $n_0$. How it has been shown in [17], for temperatures $T \sim 100$ K and carrier density $n_0 \sim 10^{11}$ cm$^{-2}$ value of chemical potential is $\mu_{chem} \sim 3.5 \cdot 10^{-14}$ erg.

### III. RESULTS AND DISCUSSION

Due to resonant dependencies of components of permittivity, permeability tensors of magnetic semiconductor (1), (2) and conductivity tensor of graphene (3), it is clear that reflectance, transmittance and absorptance will have some features near the resonant. At the frequencies near the ferromagnetic resonant frequency $\omega \cong \omega_H (1+\alpha^2)^{-1/2}$ spin oscillations (or magnons) will excite in magnetic semiconductor, and, therefore, electromagnetic wave will be strongly damped in the medium. This leads to decrease in $T$ and increase in $R$ and $A$. Near the cyclotron resonance frequency $\omega \cong \omega_c (1+\nu^2)^{-1/2}$ the energy of electromagnetic wave actively converse to energy of the electrons rotating around the magnetic field lines. Thus, electromagnetic radiation of these frequencies will also strongly absorb by medium. The resonant frequencies are determined by the value of external magnetic field. Effect of graphene layers on the electromagnetic properties of the structure is more evident at the frequencies corresponding to carriers' transitions between Landau levels. At these frequencies, conductivity of graphene sharply increases, that lead to increase in absorptance and reflectance (the structure becomes more "metallic") and to decrease in transmittance.

Fig. 2 shows the frequency dependencies of $R$, $T$, and $A$ for the structure placed in magnetic field of 20 kOe. One can see that increasing in number of periods with fixed size of one period leads to increasing in reflectance and absorptance and decreasing in transmittance. This also leads to shift in size resonance frequencies caused by interference of waves reflected from different borders of the layers. In the frequency dependence of transmittance shown in Fig. 2 there practically is no feature associated with a ferromagnetic resonance. This is due to insufficient thickness of the magnetic semiconductor layer to observe a significant effect. Increase of thickness of the structure leads to the fact, that the electromagnetic wave runs longer path in magnetic semiconductor and, consequently, more energy will be transferred to magnetic subsystem. For the structure with the fixed size, increasing in number of graphene layers from 2 to 11 (see right panel of Fig. 2) leads to increasing in reflectance and decreasing in transmittance and absorptance on about 3-4%. Decreasing in absorptance is caused by the following fact: some part of electromagnetic wave reflects from each graphene layer. So, an effective length electromagnetic wave goes into the material is less than without graphene layers in the structure. At frequencies of graphene electrons transitions between Landau levels electromagnetic power losses in graphene layers become more significant than an effect of reflection, and features associated with such transitions become more evident with increasing in number of graphene layers.

Resonant frequencies values, energy of Landau levels in graphene, and distance between them depend on the value of external magnetic field. Fig. 3 shows field dependencies of reflectance, transmittance and absorptance of electromagnetic wave by graphene - magnetic semiconductor superlattice for frequencies $f = \omega/2\pi = 1$ THz and $f = 5$ THz. The field dependencies have a resonance form. Calculation shows, that for $f < 3$-4 THz there is mainly one resonance, which is associated with the cyclotron resonance in magnetic semiconductor layers. At frequencies $f > 3$-4 THz there are two coupled resonances: the cyclotron one and one, which is associated with graphene electrons transitions between Landau levels. Increase in number of periods of the structure leads to increasing in resonant value of reflectance and to decreasing in resonant values of transmittance and absorptance for $f < 3$-4 THz. The overall scenario of the change in reflectance, transmittance and absorptance with increasing of magnetic field in this frequency range is as follows: the sharp change to resonance (increasing in $R$, decreasing in $T$ and $A$) and then slow relaxation. At frequencies $f > 3$-4 THz the slump is replaced by slow decrease in $R$ (increase in $A$) at magnetic field value about 10 kOe and take a turning-point at magnetic field about 15-20 kOe. Magnetic field value of this turning-point depends from the number of periods $N$ of the structure and goes to lower magnetic fields with increasing in $N$. this is due to interaction of carriers oscillations in graphene layers. Behavior of transmittance $T$ is the different: at low magnetic fields $T$ is increasing and take a maximum at magnetic field values about 10 kOe, then it gradually decreases to a minimum at high magnetic fields about 25 kOe. The critical magnetic field values for the transmittance behavior almost not depend on the number of periods of the structure, but increasing in $N$ make the change in transmittance to be greater.

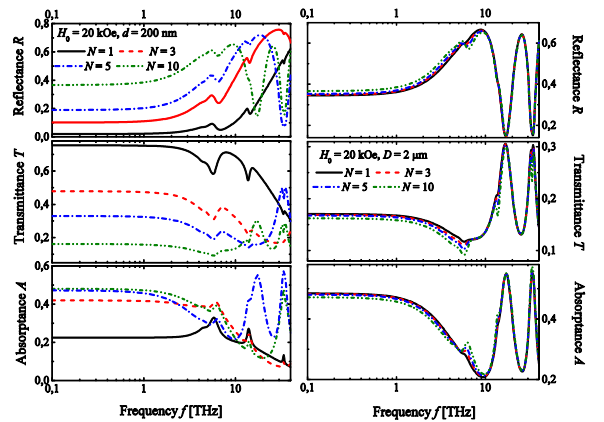

Fig. 2. (Color online) Frequency dependencies of electromagnetic waves reflectance $R$, transmittance $T$, and absorptance $A$ for different number of periods $N$. Left panel: fixed size of one period $d$; right panel: fixed size of all structure $D$. An external magnetic field value $H_0 = 20$ kOe.

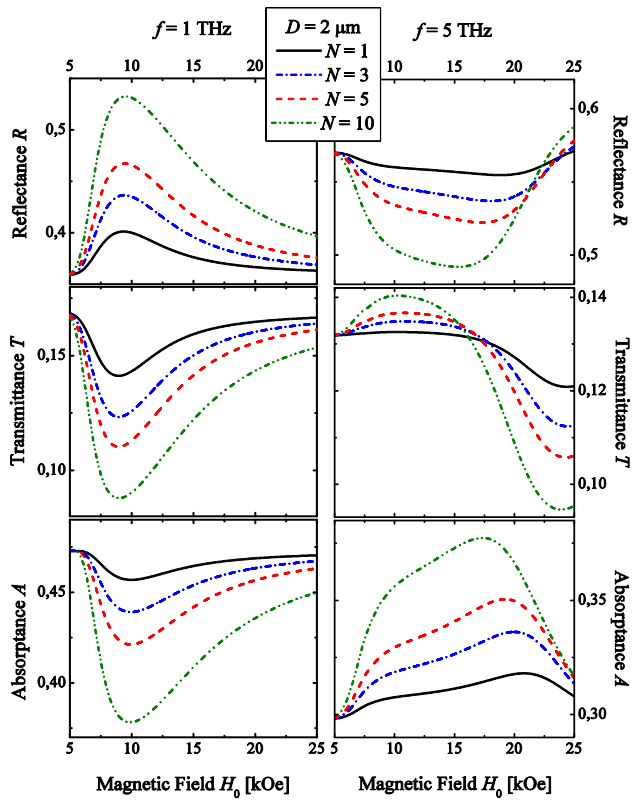

Fig. 3. (Color online) Field dependencies of electromagnetic waves reflectance $R$, transmittance $T$, and absorptance $A$ for different number of periods $N$. Left panel: $f = \omega/2\pi = 1$ THz; right panel: $f = 5$ THz. The size of all structure $D = 2$ μm.

IV. CONCLUSION

Investigation of electrodynamic characteristics of graphene - magnetic semiconductor superlattice placed in an external magnetic field showed that electrodynamic characteristics of such structure can be efficiently controlled. Reflectance, transmittance and absorptance of electromagnetic waves can be changed with the change of external magnetic field and the number of periods of the structure. At frequencies $f < 3$-4 THz, this change may reach more than 10% in relatively weak magnetic fields about 10 kOe. At $f > 3$-4 THz the change of $R$ and $A$ is not so great: a little about 5% at magnetic field values of 10 kOe and a little less than 10% at magnetic field values of 15-20 kOe. The change in $T$ at this frequency range is even less: about 1% at 10 kOe and about 5% at 25 kOe.

For room temperature applications magnetic semiconductors with higher $T_C$ should be used: for example, $Cd_{1-x}Mn_xGeP_2$ [18], $Co_xTi_{1-x}O_2$ [19] or $Zn_{1-x}Mn_xO$ [20]. But features of graphene conductivity at high temperature will not to be so sharp. So, the effects discussed in present paper at room temperature may be less then we have predicted.